\documentclass[runningheads]{llncs}

\usepackage{amsmath,amssymb,amsfonts, mathtools}
\usepackage[ruled,vlined, linesnumbered]{algorithm2e}
\usepackage{graphicx}
\usepackage{textcomp}
\usepackage{xcolor}
\usepackage{booktabs} 
\usepackage{hyperref}
\usepackage{amsmath}
\usepackage{enumitem}
\usepackage{relsize}
\usepackage{subcaption}
\usepackage{multirow}
\usepackage[depth=3]{bookmark}
\usepackage{marvosym}

\begin{document}
\title{Experimental Analysis of Machine Learning Techniques for Finding Search Radius in Locality Sensitive Hashing}
\author{Omid Jafari \Letter\orcidID{0000-0003-3422-2755} \and
Parth Nagarkar\orcidID{0000-0001-6284-9251}
}
\authorrunning{O. Jafari et al.}

\institute{New Mexico State University, Las Cruces, New Mexico, USA \\
\email{\{ojafari, nagarkar\}@nmsu.edu}}
\maketitle              
\begin{abstract}
Finding similar data in high-dimensional spaces is one of the important tasks in multimedia applications. Approaches introduced to find exact searching techniques often use tree-based index structures which are known to suffer from the \textit{curse of the dimensionality} problem that limits their performance. Approximate searching techniques prefer performance over accuracy and they return \textit{good enough} results while achieving a better performance. Locality Sensitive Hashing (LSH) is one of the most popular approximate nearest neighbor search techniques for high-dimensional spaces. One of the most time-consuming processes in LSH is to find the neighboring points in the projected spaces. An improved LSH-based index structure, called \underline{r}adius-\underline{o}ptimized \underline{L}ocality \underline{S}ensitive \underline{H}ashing (\textit{roLSH}) has been proposed to utilize Machine Learning and efficiently find these neighboring points; thus, further improve the overall performance of LSH. In this paper, we extend roLSH by experimentally studying the effect of different types of famous Machine Learning techniques on overall performance. We compare ten regression techniques on four real-world datasets and show that Neural Network-based techniques are the best fit to be used in roLSH as their accuracy and performance trade-off are the best compared to the other techniques.

\end{abstract}
\section{Introduction}
Efficient processing of nearest neighbor queries in high-dimensional spaces is an important task in many large scale multimedia retrieval applications. Exact nearest neighbor search techniques that use tree-based index structures, such as KD-tree, SR-tree, etc., work well in low-dimensional spaces ($< 10$), but they suffer from the \textit{curse of dimensionality} problem in high-dimensional spaces. The exact nearest neighbor search techniques are often outperformed by linear search techniques \cite{chavez2001searching}. Using \textit{good enough} approximate results instead of exact results is a solution to this problem. The techniques that yield approximate results sacrifice a little accuracy for a significant gain in the overall processing time. This trade-off is useful in many applications where a $100\%$ accuracy is not required. The goal of the approximate nearest neighbor search problem, also called \textit{c-approximate Nearest Neighbor search}, is to return points that are within $c \times R$ distance from the given query, where the approximation ratio $c$ is a user-defined parameter and $R$ defines the distance of the query and its nearest neighbor.

\subsection{Locality Sensitive Hashing}
One of the most popular techniques for finding approximate nearest neighbors in high-dimensional spaces is Locality Sensitive Hashing (LSH) \cite{gionis1999similarity}. It was first proposed in \cite{gionis1999similarity} for the Hamming distance but was afterward proposed for other distances such as the popular Euclidean distance \cite{datar2004locality}. LSH uses \textit{randomized} hash functions to map the data in the original high-dimensional space to several low-dimensional spaces called projections. The main idea behind LSH is that the points that are close in the original space will map to the same hash buckets in the low-dimensional spaces with a higher probability compared to the points that are far from each other in the original space. Since LSH is used in many different applications \cite{jafari2021survey}, there have been several works that have been proposed to improve its accuracy and performance \cite{lv2007multi,gan2012locality,huang2015query,christiani2019fast,jafari2019qwlsh,liu2019lsh,liu2014sk,jafari2020improving,lu2020r2lsh}.

\subsection{Motivation}
Locality Sensitive Hashing (LSH) performs in a sub-linear time complexity (in terms of the data size) and provides theoretical guarantees on the accuracy of its results. Furthermore, it is a data-independent technique that is not affected by changes in data properties such as data distribution. Hence, LSH is a useful technique in many applications where data independency and scalability are important. While the traditional LSH index structures suffered from disk I/Os, novel index structures, such as roLSH \cite{jafari2020improving}, have been proposed to improve the disk I/Os, and thus, the performance of LSH. To improve the disk I/Os (specifically, the random disk I/Os), roLSH utilizes machine learning techniques to predict the search radius before beginning query processing. Since prediction is the main part of roLSH, it is crucial to use a prediction method that not only achieves good accuracy but also in a small amount of time.

\subsection{Contributions}
In this paper, we carefully present a detailed experimental analysis of several machine learning techniques that can be used in the prediction phase of roLSH. Our contributions are as follows:

\begin{itemize}
    \item We compare the prediction techniques on real-world datasets with different characteristics and different experimental settings.
    \item We show the importance of analyzing different evaluation metrics and considering all the trade-offs involved in the comparison.
\end{itemize}


\section{Background and Preliminaries}
\label{sec:background}
In this section, we provide some background on LSH using the terminologies and formulations introduces in the traditional methods E2LSH \cite{datar2004locality} and C2LSH \cite{gan2012locality}.

\noindent \textbf{Hash Functions:} A hash function family $H$ is ($R$, $cR$, $p_1$, $p_2$)-sensitive if it satisfies the following conditions for any two points $x$ and $y$
in a $d$-dimensional dataset $D \subset \mathbb{R}^d$:

\begin{itemize}
	\item if $|x - y| \leq R$, then $Pr[H(x) = H(y)] \geq p_1$, and
	\item if $|x - y| > cR$, then $Pr[H(x) = H(y)] \leq p_2$
\end{itemize}

In the formulas above, $p_1$ and $p_2$ are probabilities and $c$ is an approximation ratio. The formulas state that points $x$ and $y$ are hashed to the same hash bucket with a high probability (greater than $p_1$) if they are close to each other in the original space, and are hashed to the same hash bucket with a low probability (less than $p_2$) if they are not close to each other in the original space. For Euclidean distance, each LSH hash function is defined as $H_{\vec{a},b} (x) = \left\lfloor{\frac{\vec{a}.x + b}{w}}\right\rfloor,$ where $\vec{a}$ is a $d$-dimensional random vector with entries chosen independently from the standard normal distribution $N(0,1)$ and $b$ is a real number chosen uniformly from $[0, w)$, such that $w$ is the width of the hash bucket \cite{datar2004locality}.

\noindent \textbf{Collision Counting:} C2LSH \cite{gan2012locality} proves that only points that are mapped to the same bucket with the query (i.e. colliding with the query) in at least $l$ projections (out of $m$) need to be chosen as candidates. Here, $l$ is the collision count threshold which is calculated based on the dataset size and user-provided parameters.

\noindent \textbf{Virtual Rehashing:} C2LSH \cite{gan2012locality} begins query processing with a very small radius, and later on, exponentially increases the radius if not enough candidates are found at the current search radius. 

\section{Problem Specification}
As mentioned in Section \ref{sec:background}, LSH methods that use Virtual Rehashing in the query processing, start the search with an initial radius of $1$ and if not enough candidates are found, then the radius is increased in an exponential fashion. roLSH \cite{jafari2020improving}, shows that the exponential increase of the radius is not efficient and leads to excessive disk I/Os; thus, roLSH proposes an improved index structure that uses sampling and machine learning techniques to predict the optimized radius for a given query. Although roLSH provides novel techniques to fix errors in the predicted radius, its performance and accuracy can be affected by the chosen machine learning technique. In this paper, our goal is to study and compare several machine learning techniques that can be used by roLSH.

\section{Experimental Analysis}
In this section, we first describe our experimental evaluation plan. We experimentally analyze different machine learning techniques (i.e. regression techniques) that can be used in roLSH. Specifically, we examine the Linear Regression, RANSAC, Ridge, Lasso, Decision Tree, Gradient Boosting, Bayesian, Elastic Net, Support Vector Regression (SVR), and Multi Layer Perceptron (MLP) regression techniques. All experiments were run on the nodes of the Discovery cluster with the following specifications on each node: Intel Xeon E5-2650, 8 CPU cores, 16GB RAM, and Red Hat Enterprise Linux (RHEL) 8.6 operating system. All codes related to roLSH were written in C++11 and compiled with GCC v4.7.2 with the -O3 optimization flag. The codes related to the regression techniques were written in Python 3.10 and using version 1.1 of the Scikit-learn Python package \cite{scikit-learn}. 

\subsection{Datasets}
We use the following four diverse high-dimensional datasets with varying cardinality and dimensionality:

\begin{itemize}
	\item \textbf{LabelMe}\cite{russell2008labelme} consists of $181,093$ 512-dimensional points which were generated by running the GIST feature extraction algorithm on $30369$ annotated images belonging to $183$ categories. There are no duplicates in the dataset and values range between zero and $58104$.
	
	\item \textbf{Sift1M}\cite{jegou2010product} consists of $1,000,000$ 128-dimensional points that were created by running the SIFT feature extraction algorithm on real images. The values of this dataset are integers between zero and $218$.
	
	\item \textbf{Deep1M} consists of $1,000,000$ 96-dimensional points sampled from the Deep1B dataset introduced in \cite{babenko2016efficient}. These points are extracted from the last layers of convolutional neural networks for images.
	
	\item \textbf{Mnist8M}\cite{loosli2007training} This dataset, also known as the InfiMNIST dataset, contains $8,100,000$ 784-dimensional points that represent images of the digits 0 to 9 which are grayscale and of size 28 $\times$ 28.
\end{itemize}

\begin{figure*}[!h]
	\centering
	{\includegraphics[width=\linewidth]{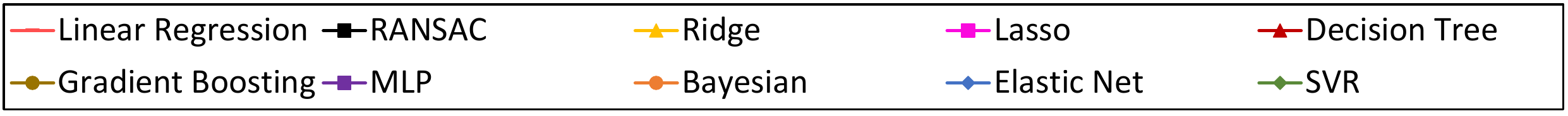}}
	
	\begin{subfigure}[b]{0.47\textwidth}
		\centering
		{\includegraphics[width=\linewidth, height=1.5in]{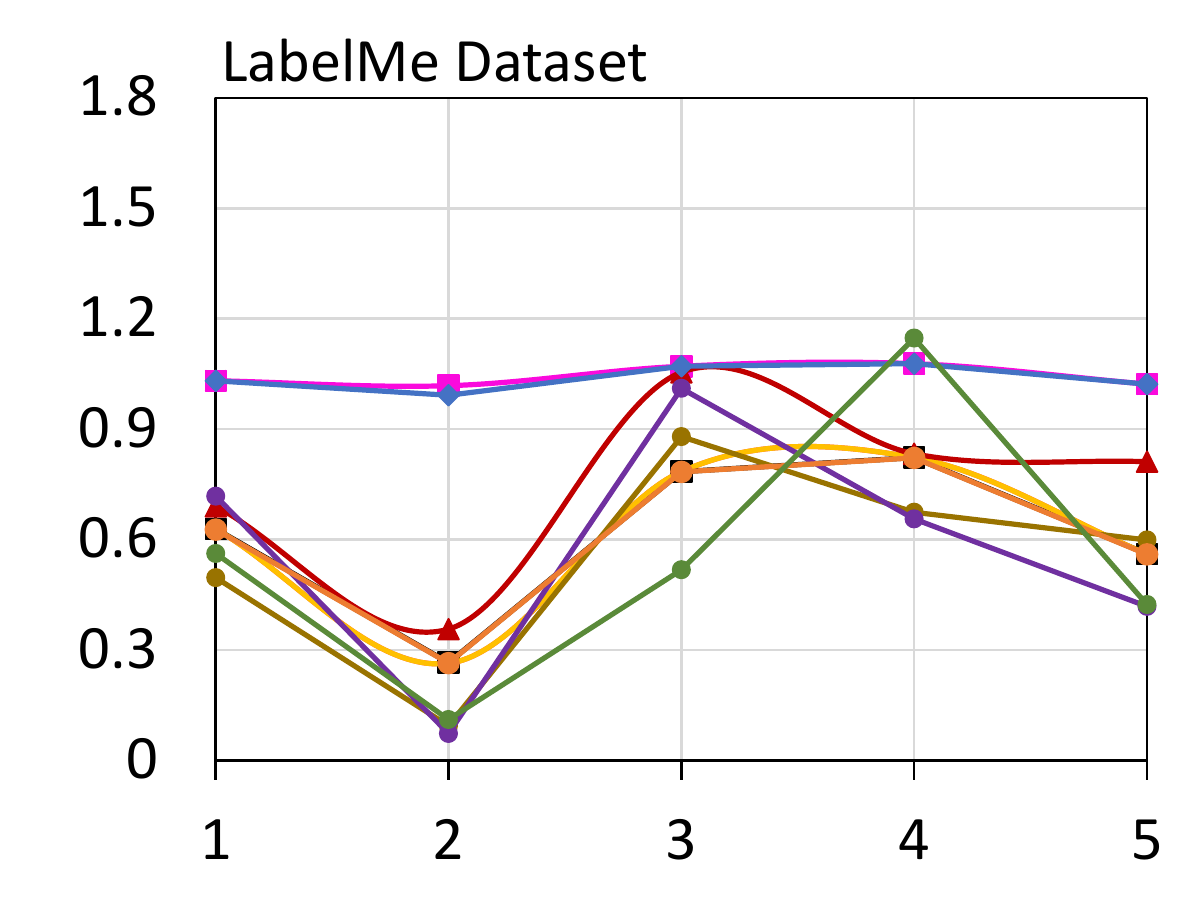}}
	\end{subfigure}\quad
	\begin{subfigure}[b]{0.47\textwidth}
		\centering
		{\includegraphics[width=\linewidth, height=1.5in]{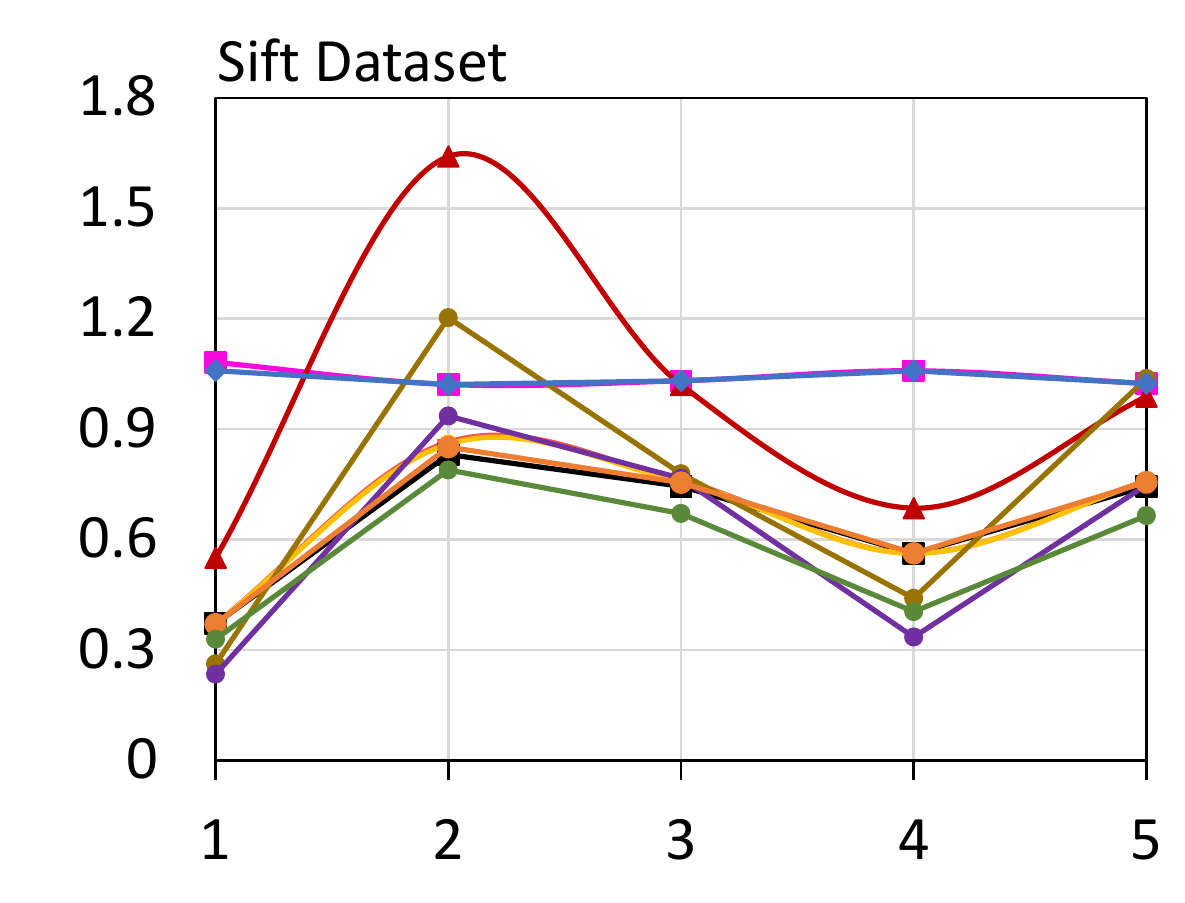}}
	\end{subfigure} \\

	\begin{subfigure}[b]{0.47\textwidth}
		\centering
		{\includegraphics[width=\linewidth, height=1.5in]{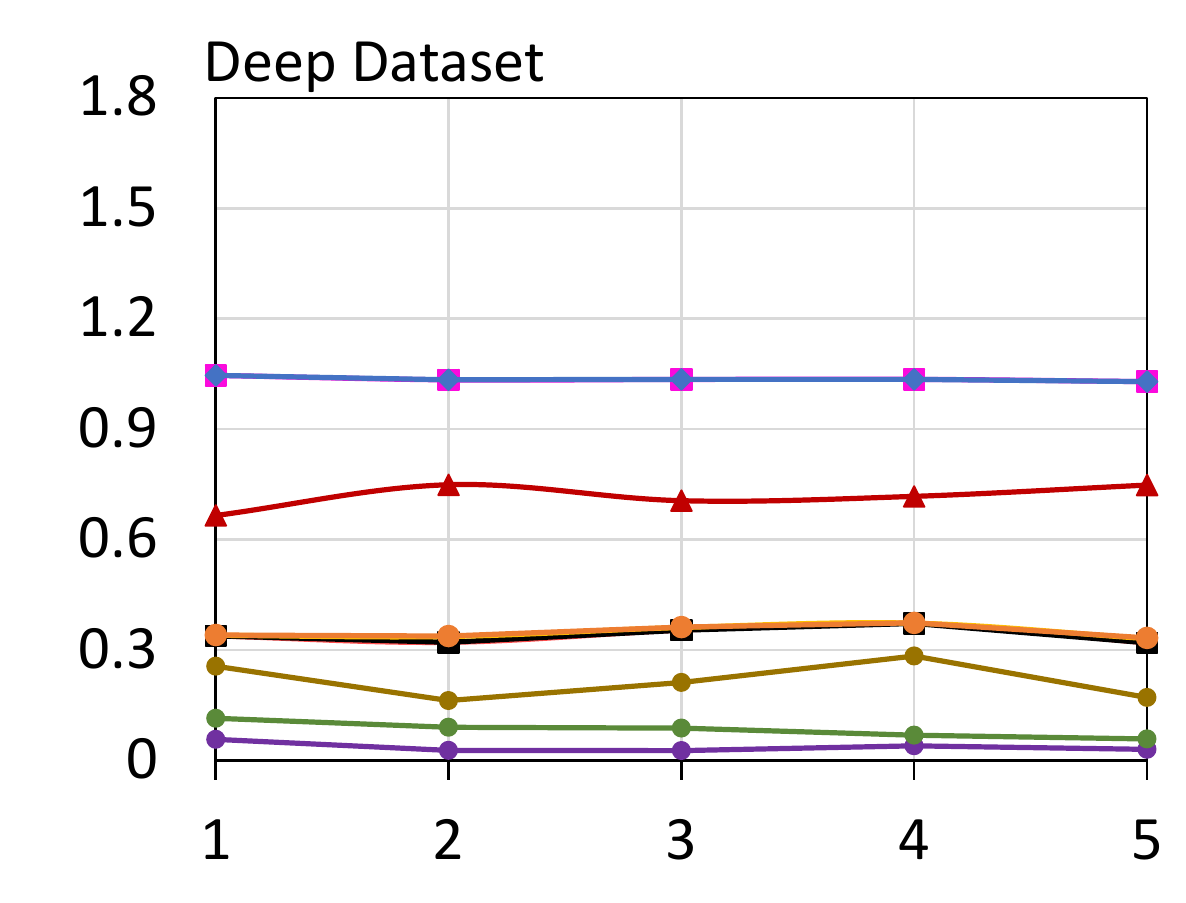}}
	\end{subfigure}\quad
	\begin{subfigure}[b]{0.47\textwidth}
		\centering
		{\includegraphics[width=\linewidth, height=1.5in]{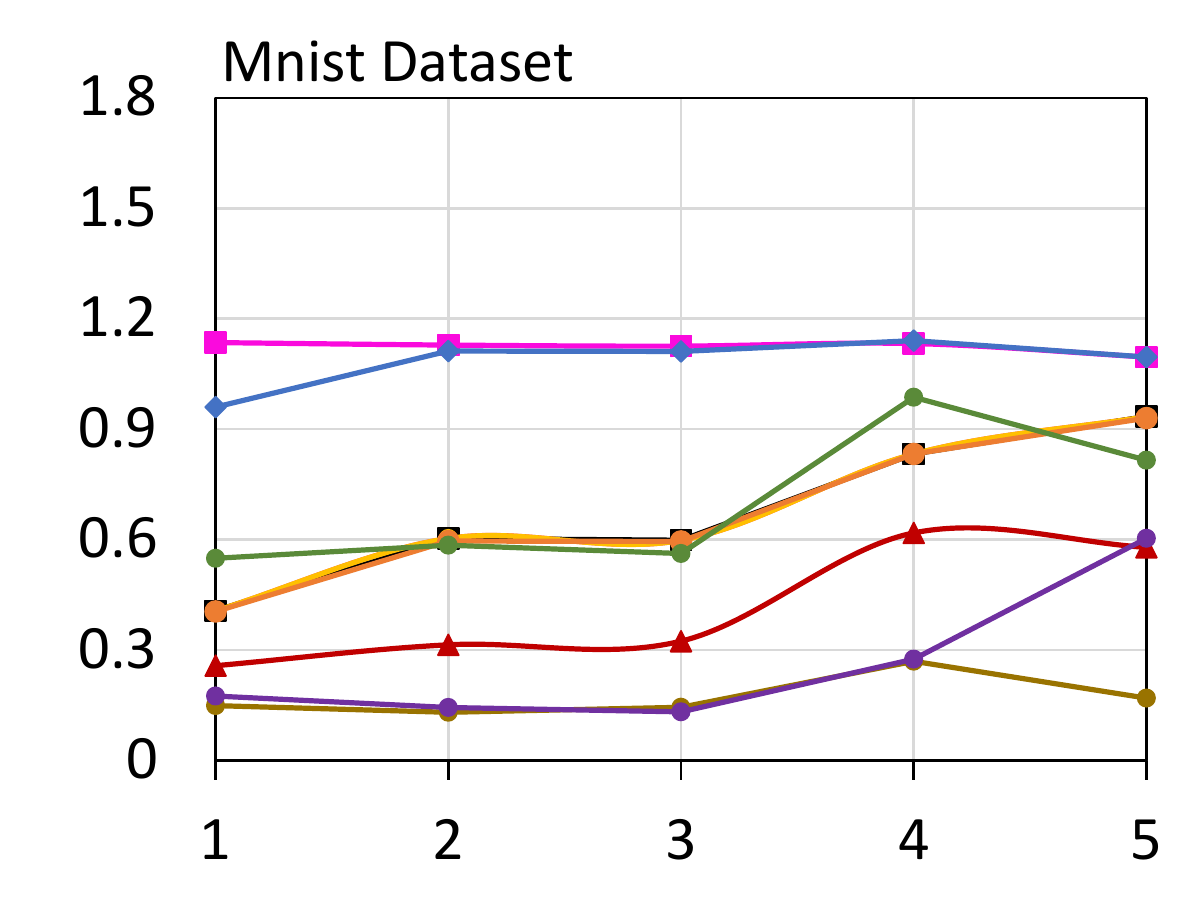}}
	\end{subfigure}

	\caption{Mean Squared Error (Y Axis) for different training scenarios (X Axis)}
	\label{fig:expMSE}

	\bigskip

	\begin{subfigure}[b]{0.47\textwidth}
		\centering
		{\includegraphics[width=\linewidth, height=1.5in]{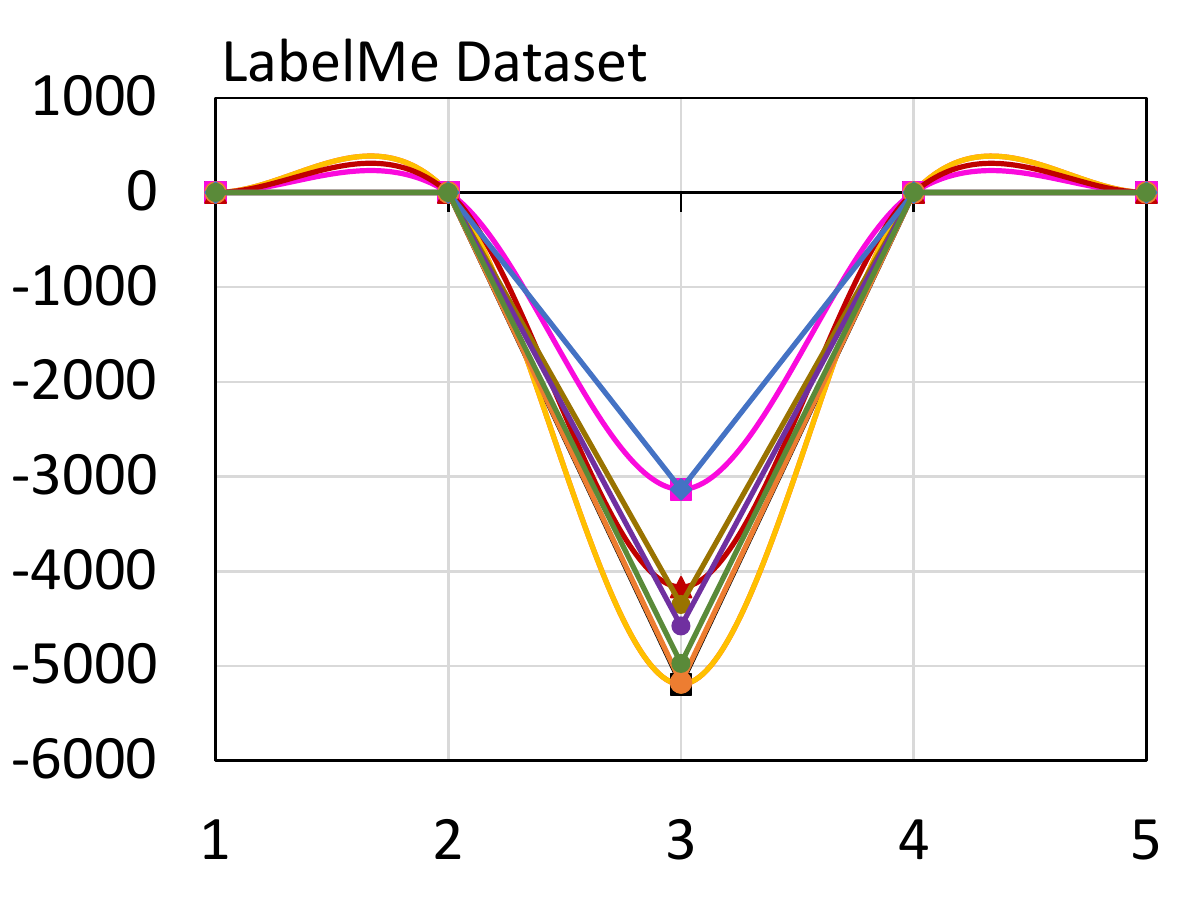}}
	\end{subfigure}\quad
	\begin{subfigure}[b]{0.47\textwidth}
		\centering
		{\includegraphics[width=\linewidth, height=1.5in]{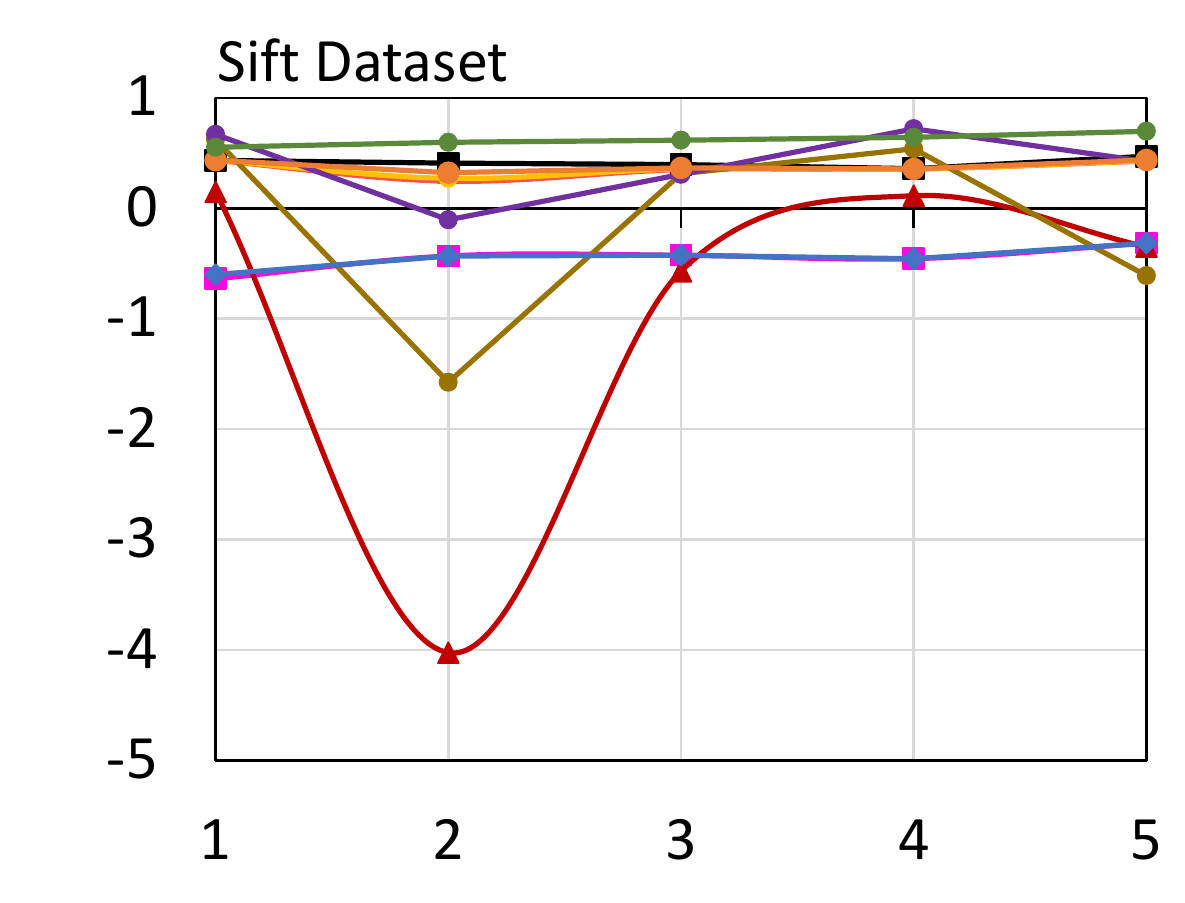}}
	\end{subfigure} \\

	\begin{subfigure}[b]{0.47\textwidth}
		\centering
		{\includegraphics[width=\linewidth, height=1.5in]{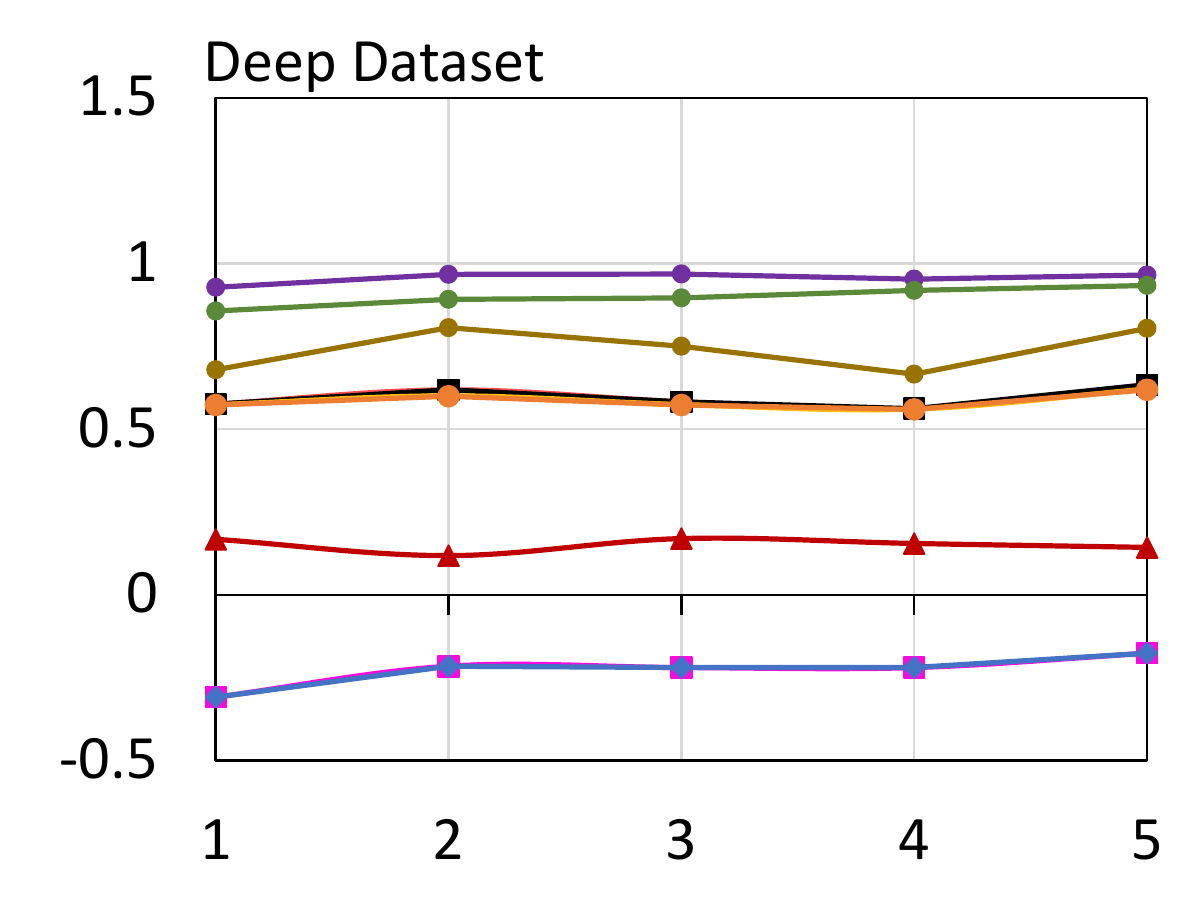}}
	\end{subfigure}\quad
	\begin{subfigure}[b]{0.47\textwidth}
		\centering
		{\includegraphics[width=\linewidth, height=1.5in]{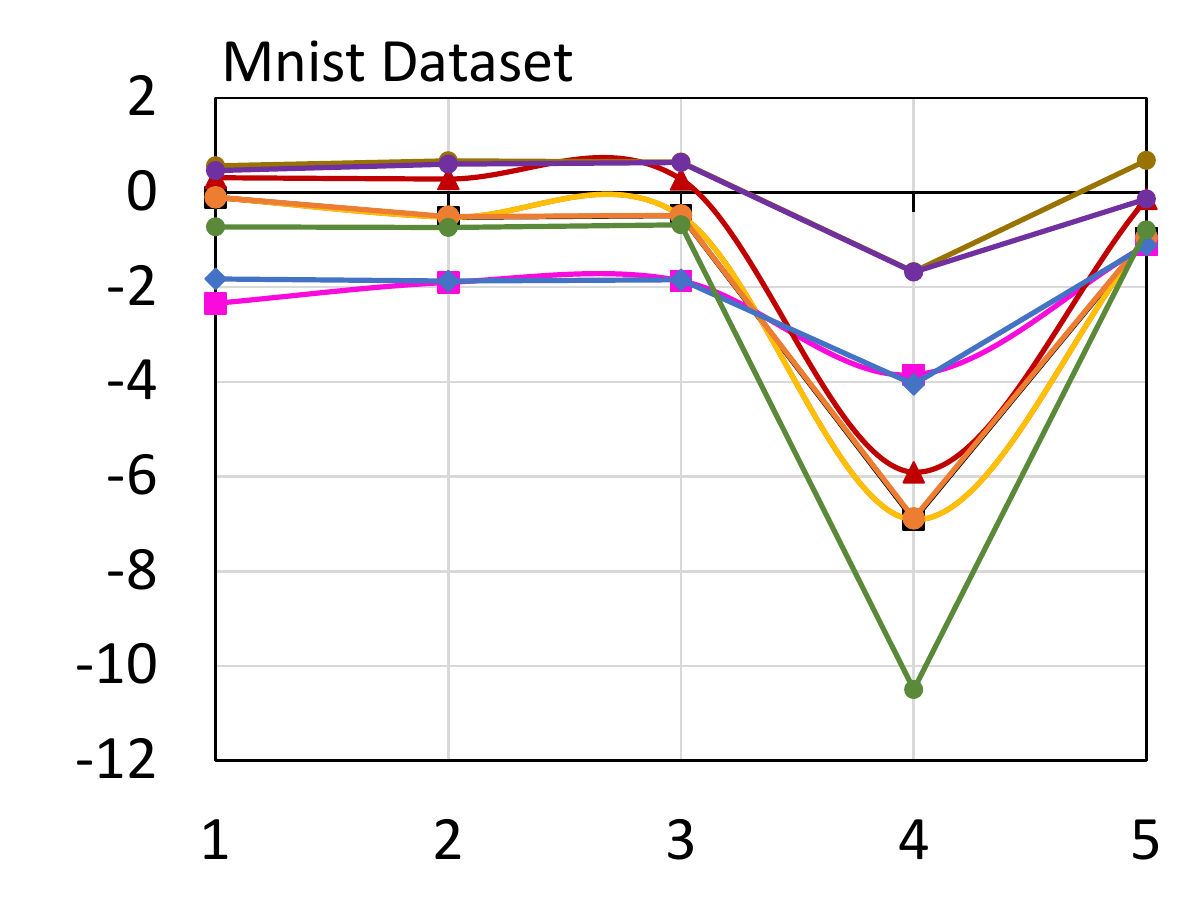}}
	\end{subfigure}

	\caption{R-Squared (Y Axis) for different training scenarios (X Axis)}
	\label{fig:expR2}
\end{figure*}

\subsection{Evaluation Criteria and Parameters}
The goal of our paper is to present an experimental study of the effect of different types of Machine Learning techniques on the performance of roLSH. The performance and accuracy of the compared techniques used in this paper are evaluated using the following metrics:

\begin{itemize}
    \item \textbf{Mean Squared Error (MSE):} MSE shows the average of the squared distances between actual and predicted values. MSE is always a positive number and is equivalent to the error (i.e. MSE of zero shows that there is no error). MSE is calculated using the following formula:
    \begin{equation}
        MSE = \frac{1}{n} \sum_{i=1}^{n}(y_i-\hat{y}_i)^2
    \end{equation}
    where $n$ is the number of evaluation data, $y$ is the actual value, and $\hat{y}$ is the predicted value.

    \item \textbf{R Squared ($R^2$):} In general, $R^2$ shows how close the data are to the fitted line by the regression model. This metric is considered the normalized version of MSE and is useful since it does not depend on the scale of the data. $R^2$ is calculated using the following formula:
    \begin{equation}
        R^2 = \frac{\sum_{i=1}^{n}(\hat{y}_i - \Bar{y})^2}{\sum_{i=1}^{n}(y_i - \Bar{y})^2}
    \end{equation}
    where $\bar{y}$ is the mean of all actual values.
    
    \item \textbf{Prediction Time:} The time it takes to predict the values for a given data.
\end{itemize}

To generate the actual values (i.e. ground truth), we use the C2LSH algorithm which is also used in roLSH. We use the parameters suggested in the C2LSH paper. Specifically, we set the bucket width ($w$) to $2.184$ and the allowed error probability ($\delta$) to $0.1$. In this paper, we use the default parameters and options of the regression techniques provided by the Python library. The parameter tuning analysis is left for future work since the default parameters yield good enough results. The testing data size is $20\%$ of the training data size which is selected randomly. Additionally, we used 10-fold cross validation in our experiments.

\begin{figure*}[!h]
	\centering
	{\includegraphics[width=\linewidth]{Figures/Legend.pdf}}
	
	\begin{subfigure}[b]{0.47\textwidth}
		\centering
		{\includegraphics[width=\linewidth, height=1.5in]{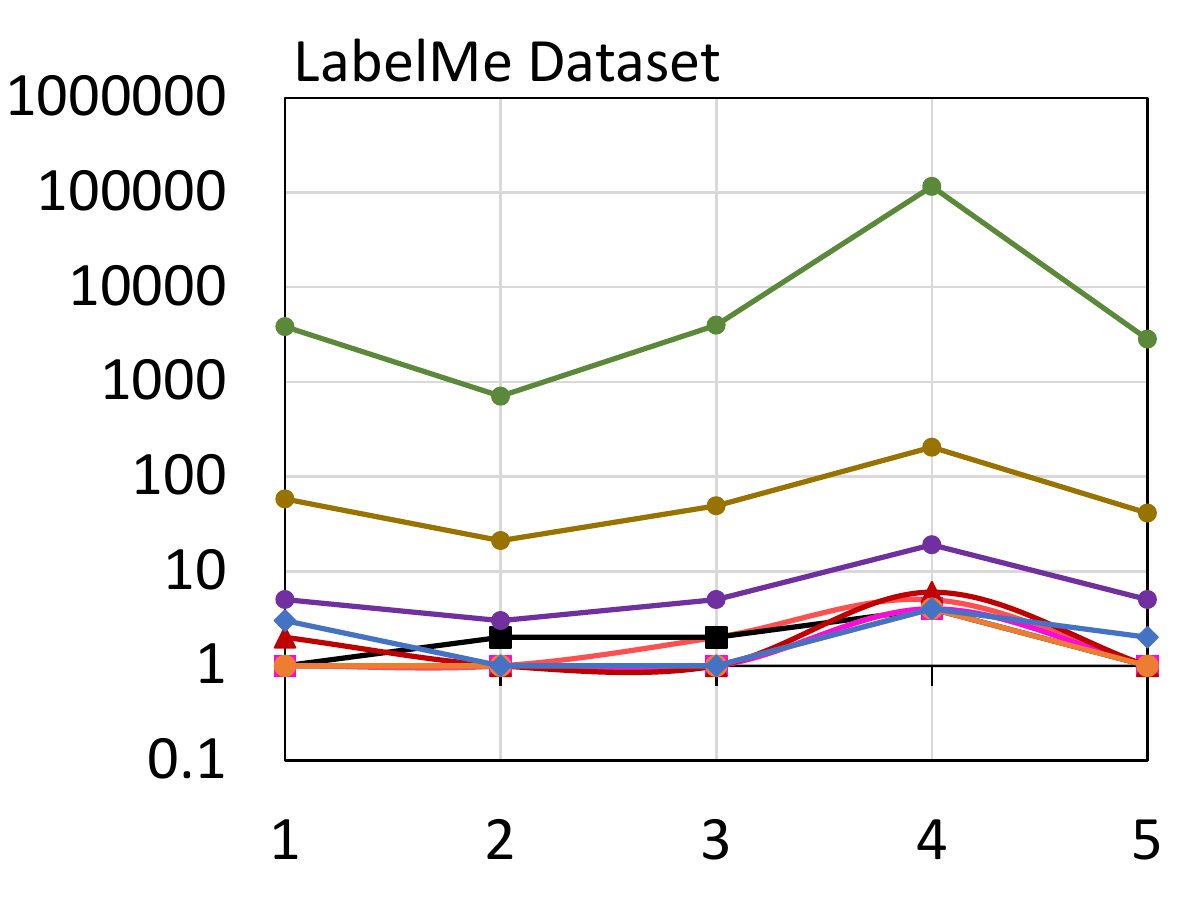}}
	\end{subfigure}\quad
	\begin{subfigure}[b]{0.47\textwidth}
		\centering
		{\includegraphics[width=\linewidth, height=1.5in]{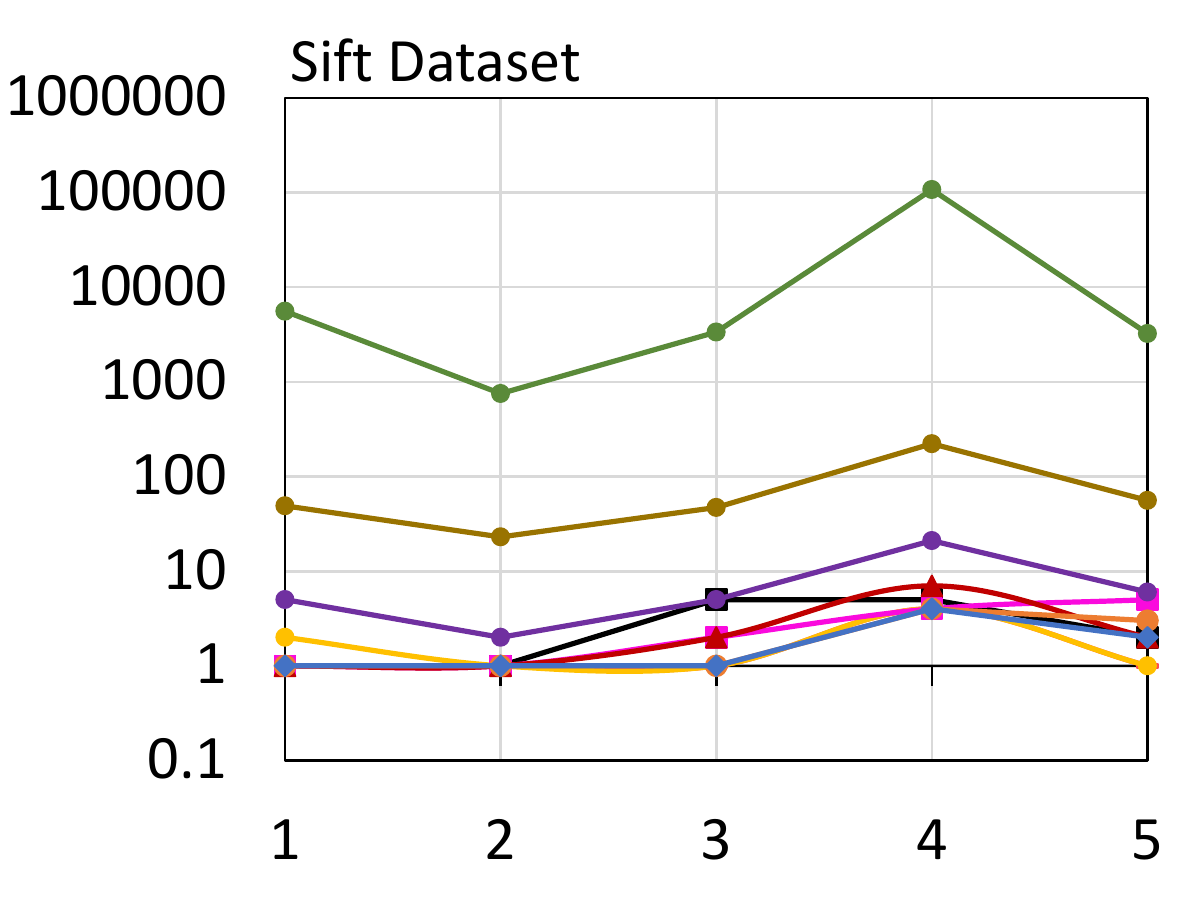}}
	\end{subfigure} \\

	\begin{subfigure}[b]{0.47\textwidth}
		\centering
		{\includegraphics[width=\linewidth, height=1.5in]{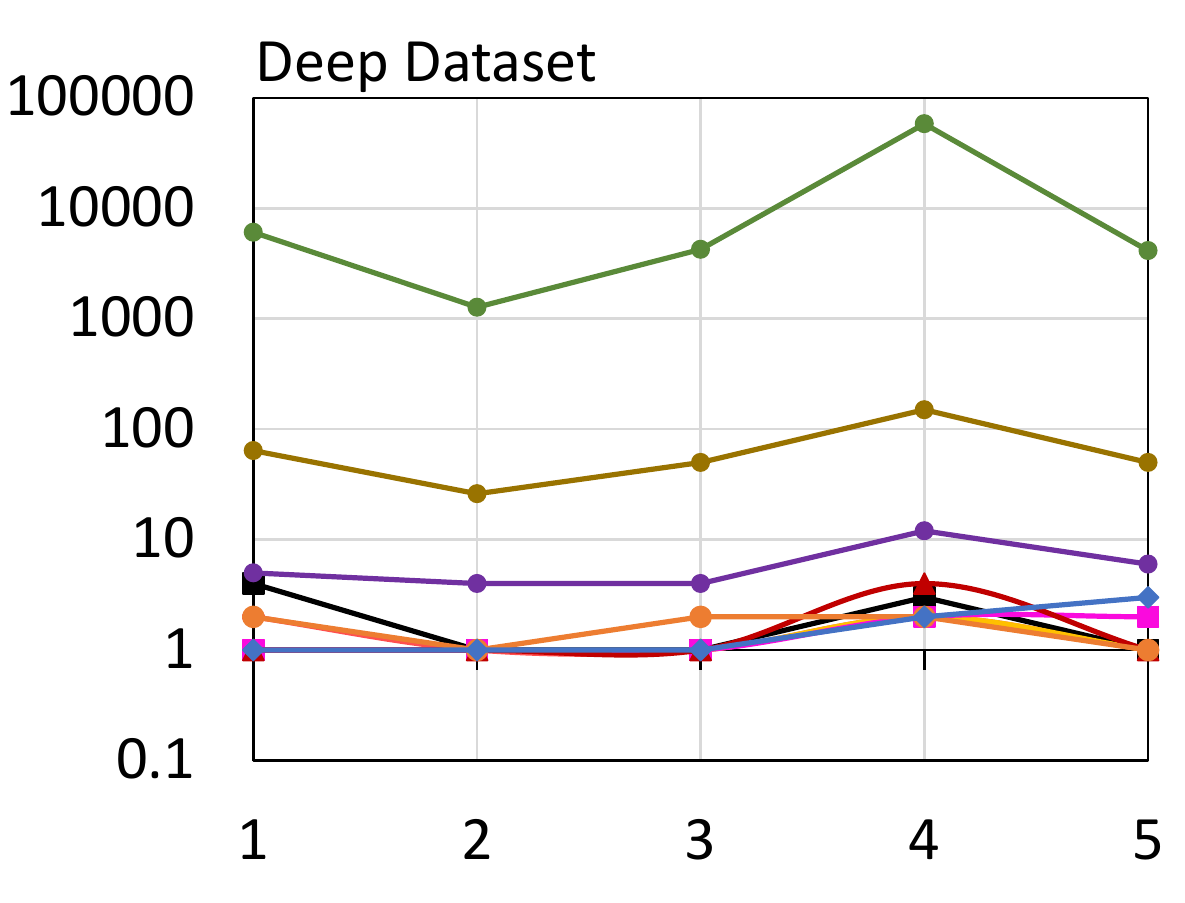}}
	\end{subfigure}\quad
	\begin{subfigure}[b]{0.47\textwidth}
		\centering
		{\includegraphics[width=\linewidth, height=1.5in]{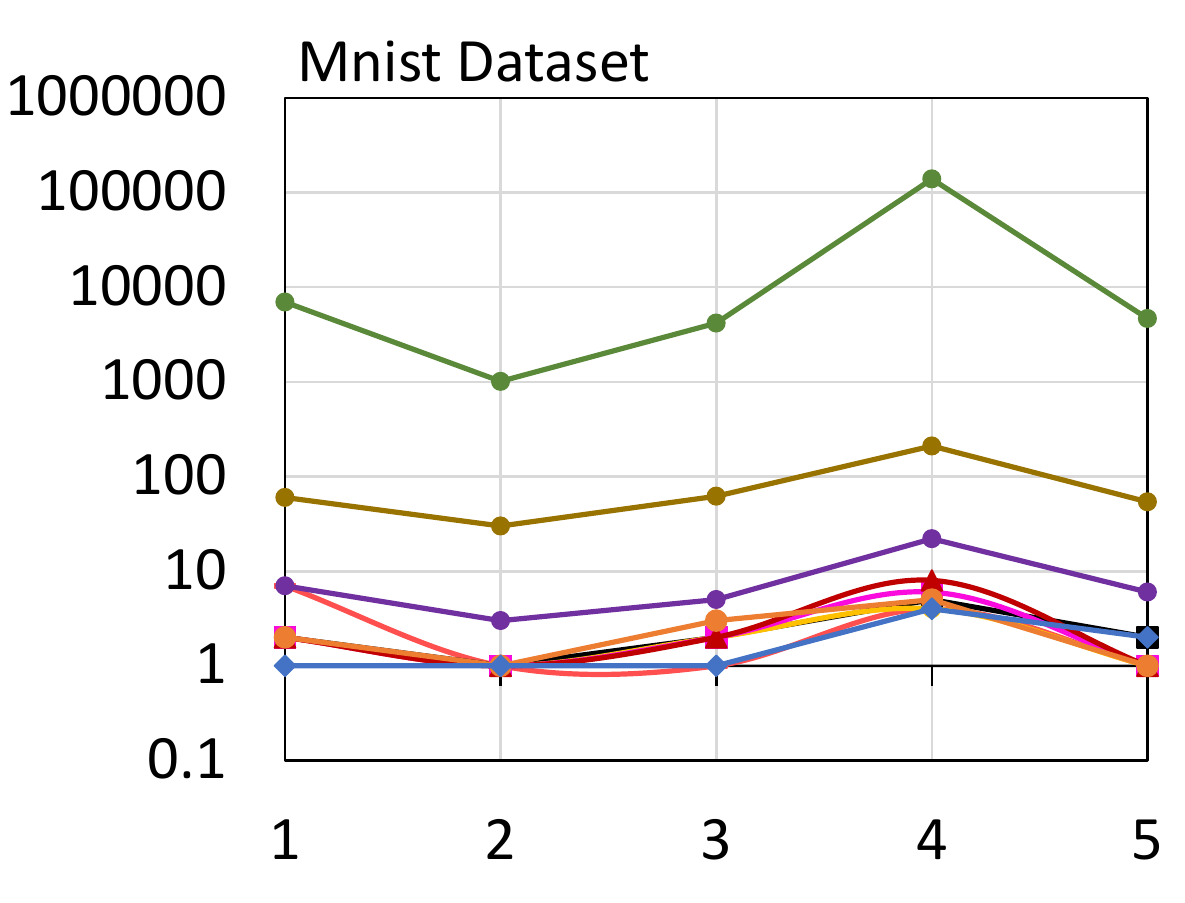}}
	\end{subfigure}

	\caption{Prediction Time (in ms, \textit{log scale}) (Y Axis) for different training scenarios (X Axis)}
	\label{fig:expTime}
\end{figure*}

\subsection{Training Scenarios}
As prediction techniques depend greatly on the data they are trained on, we used different training scenarios in our experiments. These scenarios are as follows:

\begin{enumerate}
    \item In this scenario, the training data contains only values related to $k = 1$, $50$, and $100$. Additionally, the total training size is $10,000$.
    \item In this scenario, the training data contains only values related to $k = 1$, $25$, $50$, $75$, and $100$. Additionally, the total training size is $5,000$.
    \item In this scenario, the training data contains only values related to $k = 1$, $25$, $50$, $75$, and $100$. Additionally, the total training size is $10,000$.
    \item In this scenario, the training data contains only values related to $k = 1$, $25$, $50$, $75$, and $100$. Additionally, the total training size is $50,000$.
    \item In this scenario, the training data contains only values related to $k = 1$, $10$, $25$, $50$, $75$, $90$, and $100$. Additionally, the total training size is $10,000$.
\end{enumerate}

\subsection{Discussion of Results}
\noindent \textbf{Mean Squared Error (MSE):}
Figure \ref{fig:expMSE} shows the Mean Squared Error of the experimented techniques. For the Deep and Mnist datasets, Multi Layer Perceptron (MLP) has the lowest MSE. However, for the Sift and LabelMe datasets, Support Vector Regression (SVR) has the lowest MSE followed by MLP. The interesting observation is that the MSE of Deep and Mnist datasets are not greatly affected throughout the different training scenarios. On the other hand, LabelMe and Sift datasets show unpredictable behavior in terms of the training scenarios. This issue can be happening due to anomalies in the datasets. \\

\noindent \textbf{R-Squared (R2):}
Figure \ref{fig:expR2} shows the R-squared metric of the experimented techniques. A high R2 value can be translated into a better performing model. The figure shows that MLP has the highest R2 value for Deep and Mnist datasets, while SVR (followed by MLP) has the highest R2 value for Sift and LabelMe datasets. The observation that we mentioned before can also be seen in this figure. Our different training scenarios are greatly affecting the accuracy of the prediction techniques on LabelMe and Sift datasets. \\

\noindent \textbf{Prediction Time:}
Figure \ref{fig:expTime} shows the prediction time of the experimented techniques in milliseconds. The prediction times are log scaled because of the large difference between the values. SVR (followed by Gradient Boosting) has the highest prediction time for all datasets, while the other techniques have about the same prediction time. We can also observe that there is a direct correlation between training size and prediction time on all datasets.

\section{Conclusion}
One of the most popular techniques for Approximate Nearest Neighbor search in high-dimensional spaces is Locality Sensitive Hashing. radius-optimized Locality Sensitive Hashing (roLSH) is a unique LSH-based index structure that utilizes machine learning to improve the efficiency of traditional LSH-based methods. The benefits of roLSH over the traditional LSH-based methods have already been shown on real-world datasets. In this paper, we further analyzed the performance of roLSH by experimentally studying different types of machine learning techniques that can be used in roLSH. For this analysis, we used various sizes of datasets and different experimental settings. The results showed us that although some techniques can achieve better accuracy, their execution time is high and those techniques are not scalable. Thus, a trade-off of accuracy and performance should always be considered when comparing the techniques. Additionally, we observed that Multi Layer Perceptron (MLP) which is a Neural Network-based technique works well in all of our experiments. It is also worth mentioning that roLSH uses MLP by default. In the future, we plan to analyze the effect of changing the default parameters on the performance and accuracy of different techniques.

\subsection*{Acknowledgements} 
This work utilized resources from the New Mexico State University High Performance Computing Group, which is directly supported by the National Science Foundation (OAC-2019000), the Student Technology Advisory Committee, and New Mexico State University and benefits from inclusion in various grants (DoD ARO-W911NF1810454; NSF EPSCoR OIA-1757207; Partnership for the Advancement of Cancer Research, supported in part by NCI grants U54 CA132383 (NMSU)).

\bibliographystyle{splncs04}
\bibliography{references}

\end{document}